\documentclass{article}
\usepackage{nopageno,rmfbib,multicol,epsf,times,amsmath,amssymb,cite}
\usepackage[latin1]{inputenc}
\usepackage[]{caption2}
\usepackage{graphics}
\usepackage{graphicx}
\usepackage{multirow}
\usepackage[left=2.5cm,right=2.5cm]{geometry}

\title{Testing the molecular nature of $\phi(2170)$}
\author{A. Mart\'inez Torres${}^1$, Brenda B. Malabarba${}^1$, Xiu-Lei Ren${}^2$, K. P. Khemchandani${}^3$\\
{\small\it ${}^1$Universidade de Sao Paulo, Instituto de Fisica, C.P. 05389-970, Sao Paulo, Brazil.}\\
{\small\it ${}^2$Institut f\"ur Kernphysik $\&$ Cluster of Excelence PRISMA${}^+$, Johannes Gutenberg-Universit\"at Mainz, D-55099 Mainz, Germany.}\\
{\small\it ${}^3$Universidade Federal de Sao Paulo, C.P. 01302-907, Sao Paulo, Brazil.}}

\begin{document}

\maketitle

\begin{abstract}
\vspace{1em} In this talk we show our recent results on the decay widths of $\phi(2170)$ to final states formed by an anti-Kaon and a Kaonic resonance, in particular, $K(1460)$, $K_1(1270)$ and $K_1(1400)$, considering a molecular description for $\phi(2170)$. Branching fraction ratios are  obtained and compared with the recent results found by the BESIII collaboration, finding compatible results.\vspace{1em}
\end{abstract}

The $\phi(2170)$ has been observed by different collaborations in the last 14 years since its discovery in processes like $e^+e^-\to K^+K^-\pi^{+(0)}\pi^{-(0)}$, $J/\psi\to\eta K^+K^-\pi^+\pi^-$, $e^+e^-\to\phi\eta^\prime$ with a mass and width given by $M=2160\pm 80$ MeV and $\Gamma=125\pm 65$ MeV, respectively~\cite{BaBar:2006gsq,BaBar:2007ptr,BES:2007sqy,BESIII:2020gnc}. During this time, different quark models have been formulated to understand the nature and properties of this state, considering $\phi(2170)$ to be a $n^{2S+1}L_J=3^{3}S_1$ $s\bar s$ system, a $2^3 D_1$ $s\bar s$ system, a $s\bar s g$ hybrid system, or a tetraquark~\cite{Barnes:2002mu,Ding:2007pc,Wang:2006ri}. However such descriptions have difficulties in either finding a compatible mass and width for the state or in finding decay widths for decay channels like $K^*(892)\bar K^*(892)$, $K^*(1410)\bar K$, $K(1460)\bar K$, $K_1(1400)\bar K$ and $K_1(1270)\bar K$ compatible with the recent results of the BESIII collaboration~\cite{BESIII:2020vtu}. 

One of the interesting facts of $\phi(2170)$ is its large coupling to a $K^+K^-\pi^{+(0)}\pi^{-(0)}$ final state where $K^+K^-$ comes from the decay of $\phi$ and $\pi^{+(0)}\pi^{-(0)}$ from the decay of $f_0(980)$~\cite{BaBar:2006gsq,BaBar:2007ptr}. Indeed, the study of the $\phi K\bar K$ system and coupled channels performed in Ref.~\cite{MartinezTorres:2008gy} shows the formation of a $\phi$ meson with mass and width compatible with those of $\phi(2170)$ when the $K\bar K$ system generates $f_0(980)$. Within such a description, the cross section for the $e^+e^-\to\phi f_0(980)$ process was reproduced~\cite{MartinezTorres:2008gy,Malabarba:2020grf}. It would be then interesting to know the prediction for the decay widths of $\phi(2170)$ to a $\bar K \mathbb{K}$ system, where $\mathbb{K}$ denotes a Kaonic resonance, within the above mentioned molecular nature for $\phi(2170)$.

To do this, we require a theoretical description for $K_1(1270)$, $K_1(1400)$ and $K(1460)$ too. In case of $K(1460)$ we consider it to be generated from the interaction of $KK\bar K$ and coupled channels when one of the $K\bar K$ pairs generates $f_0(980)$~\cite{MartinezTorres:2011gjk,Albaladejo:2010tj,Filikhin:2020ksv,Zhang:2021hcl}. For $K_1(1270)$ and $K_1(1400)$ we have used different approaches: 
\begin{itemize}
\item  Model $A$: $K_1(1270)$ is considered to be a state generated from the pseudoscalar-vector interaction~\cite{Roca:2005nm,Geng:2006yb}. In this case, a two pole structure is found for $K_1(1270)$, with one of the poles being at $z_1=M-i\Gamma/2=1195-i 123$ MeV and the other at $z_2=1284-i 73$ MeV. Generation of $K_1(1400)$ is not obtained.

\item Model $B$: $K_1(1270)$ and $K_1(1400)$ are mixings of the states $K_{1A}$ and $K_{1B}$ belonging to the axial nonet, where mixing angles between $29-62^\circ$ seem to be compatible with the experimental data on their decays~\cite{Palomar:2003rb}.

\item Model $C$: We adopt a phenomenological approach, where the data available on the radiative decays of $K_1(1270)$ and $K_1(1400)$ are used to determine their couplings to the meson-meson channels involved in the decay of $\phi(2170)$.
\end{itemize}

\begin{figure}[h!]
\centering
\includegraphics[width=0.49\textwidth]{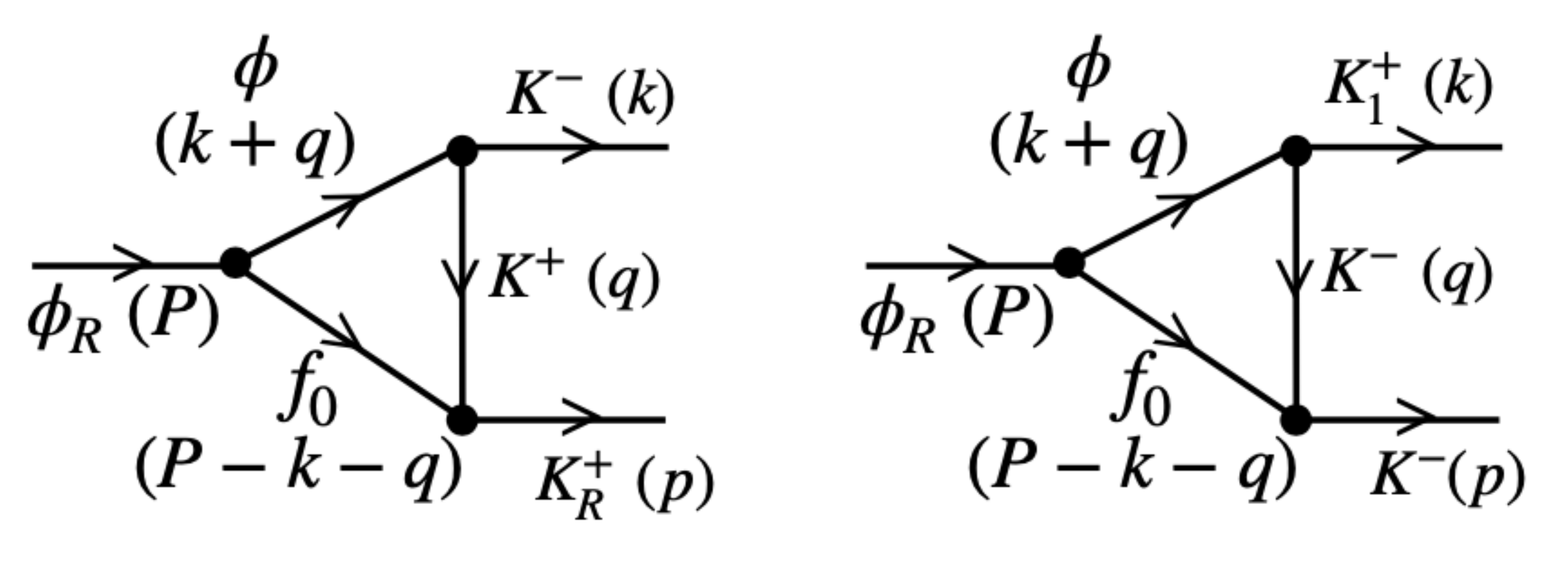}
\caption{Triangular loops involved in the process $\phi(2170)\to \bar K\mathbb{K}$. The symbols $\phi_R$, $K_R$ and $K_1$ have been used to denote $\phi(2170)$, $K(1460)$ and either $K_1(1270)$ or $K_1(1400)$, respectively. The four-momenta of the particles are written in brackets.}\label{decay}
\end{figure}

In all the preceding models, $f_0(980)$ is considered to be generated from the $K\bar K$ and $\pi\pi$ dynamics in isospin 0~\cite{Oller:1997ti}. Considering $\phi(2170)$ to be a molecular $\phi f_0(980)$ state, its decay to $\bar K\mathbb{K}$ proceeds through triangular loops (see Fig.~\ref{decay}), involving in this way three vertices: (1) First $\phi(2170)$ decays to $\phi$ and $f_0(980)$, then a $K^+$ ($K^-$) is exchanged between the $\phi$ and the $f_0(980)$, producing in this way a $f_0 K^{+} (K^-)\to K^+(1460) (K^-)$ vertex and a $\phi\to K^- (K^+_1) K^+ (K^-)$ vertex, respectively. Since the vertices $\phi(2170)\to \phi f_0(980)$, $f_0 K^+\to K^+(1460)$, $\phi\to K^+_1 K^-$ involve all of them $s$-wave interactions, we can describe them through the amplitudes
\begin{align}
t_{\phi_R}&=g_{\phi_R\to\phi f_0}\epsilon_{\phi_R}\cdot \epsilon_\phi,\nonumber\\
t_{K_R}&=g_{K^+_R\to K^+ f_0},\nonumber\\
t_{f_0\to K^+ K^-}&=g_{f_0\to K^+K^-},\nonumber\\
t_{K^+_1\to\phi K^+}&=g_{K^+_1\to \phi K}\epsilon_{K^+_1}\cdot\epsilon_\phi,\label{ts}
\end{align}
where $g_{i\to j}$ represents the coupling for the process $i\to j$ and $\epsilon_k$ is the corresponding polarization vector for particle $k$. The $\phi\to K\bar K$ vertex is described by the Lagrangian~\cite{Bando:1987br}
\begin{align}
\mathcal{L}_{VVP}=-ig \langle V^\mu[P,\partial_\mu P]\rangle,
\end{align}
with $V^\mu$ and $P$ being matrices having as elements the vector and pseudoscalar meson octet fields, respectively, $g=M_V/(2f_\pi)$, $M_V\simeq M_\rho$, $f_\pi\simeq 93$ MeV, and $\langle\quad\rangle$ indicating the SU(3) trace.

In case of models $A$ and $B$, the couplings involved in Eq.~(\ref{ts}) are obtained from the respective theoretical models used to generate $\phi(2170)$, $K(1460)$, $f_0(980)$ and $K_1(1270)$. In case of model $C$, the $K_1\to \phi K$ coupling is obtained from the radiative decay width of $K_1$ to $\gamma K^0$~\cite{ParticleDataGroup:2020ssz}. The latter decay width is calculated considering the vector meson dominance mechanism, where $\gamma$ couples to $\rho^0$, $\omega$ and $\phi$, having in this way a two step process: $K^0_1\to \rho^0 K^0+\omega K^0+\phi K^0\to \gamma K^0$. The problem within this latter approach is that we can only calculate the modulus of the coupling of $K_1\to \phi K$, and the available data allow different scenarios for this coupling in case of $K_1(1270)$. Table~\ref{coup} summarizes the couplings used in each model.
\begin{table}[h!]
\caption{Couplings (in MeV) used for the different vertices involved in Fig.~\ref{decay}.}\label{coup}
\begin{tabular}{ccccccc}
\hline
&\multicolumn{2}{c}{Model A}&Model B&\multicolumn{3}{c}{Model C}\\
\hline
&$z_1$&$z_2$&&$\mathbb{S}_1$&$\mathbb{S}_2$&$\mathbb{S}_3$\\
\hline\hline\\
$K^+_1(1270)\to \phi K^+$&$2096-i1208$&$1166-i774$&$1104\pm 171$&$3967\pm 419$&$12577\pm 763$&$19841\pm 1177$\\
$K^+_1(1400)\to\phi K^+$&\multicolumn{2}{c}{$-$}&$3533\pm 21$&\multicolumn{3}{c}{$8480\pm1333$}\\
$K^+(1460)\to K^+ f_0(980)$&\multicolumn{6}{c}{$4858\pm1337$}\\
$\phi(2170)\to \phi f_0(980)$&\multicolumn{6}{c}{$3123\pm561$}\\
\hline
\end{tabular}
\end{table}

Using the couplings listed in Table~\ref{coup} and the amplitudes in Eq.~(\ref{ts}), we can determine the amplitudes for the processes depicted in Fig.~\ref{decay} and calculate the corresponding decay widths. Indeed, the decay width for the processes shown in Fig.~\ref{decay} can be obtained as
\begin{align}
\Gamma_{i\to j}=\int d\Omega \frac{|\vec{p}^{~\text{CM}}_{i\to j}|}{32\pi^2 M^2_{\phi(2170)}}\overline{\sum\limits_\text{pol}}|t_{i\to j}|^2,\label{wphiR}
\end{align}
where $|\vec{p}^{~\text{CM}}_{i\to j}|$ is the center-of-mass momentum of the particles in the final state for the process $i\to j$, the symbol $\overline{\sum\limits_{\text{pol}}}$ indicates sum over the polarizations of the particles in the initial and final states, and average over the polarizations of the particles in the initial state, $\int d\Omega$ represents the solid angle integration and $t_{i\to j}$ is the amplitude for each of the processes depicted in Fig.~\ref{decay}.

Considering the Feynman rules, and summing over the polarizations of the internal vector mesons, we can obtained the amplitude $t_{i\to j}$. In case of the process $\phi(2170)\to K^+(1460) K^-$, we get 
\begin{align}
-it_{\phi_R\to K^+_R K^-}=&g_{\phi_R\to\phi f_0}gg_{K^+_R\to K^+f_0}\epsilon^\mu_{\phi_R}(P)\nonumber\\
&\quad\times\Bigg[k_\mu\left(1-\frac{k^2}{M^2_\phi}\right)I^{(0)}-I^{(1)}_\mu\left(1+\frac{k^2}{M^2_\phi}\right)+\frac{k_\mu}{M^2_\phi}I^{(2)}+\frac{I^{(3)}_\mu}{M^2_\phi}\Bigg],\label{tKR}
\end{align}
where
\begin{align}
I^{(0)};I^{(1)}_\mu;I^{(2)}_{\mu\nu};I^{(2)}&\equiv\int\frac{d^4q}{(2\pi)^4}\frac{1;q_\mu;q_\mu q_\nu;q^2}{\mathcal{D}},\label{Int}
\end{align}
and 
\begin{align}
\mathcal{D}=[(k+q)^2-M^2_\phi+i\epsilon][(P-k-q)^2-M^2_{f_0}+i\epsilon][q^2-M^2_K+i\epsilon].
\end{align}
Similarly, for $\phi(2170)\to K^+_1 K^-$, 
\begin{align}
-it_{\phi_R\to K^+_1 K^-}&=g_{\phi_R\to\phi f_0}g_{K^+_1\to\phi K^+}g_{f_0\to K^+ K^-}\epsilon^\mu_{\phi_R}(P)\epsilon^\nu_{K^+_1}(k)\nonumber\\
&\quad\times\Bigg[-g_{\mu\nu}I^{(0)}+\frac{k_\mu}{M^2_\phi}I^{(1)}_\nu+\frac{I^{(2)}_{\mu\nu}}{M^2_\phi}\Bigg],\label{tK1}
\end{align}
where $K^+_1$ represents either $K^+_1(1270)$ or $K^+_1(1400)$. We refer the reader to Ref.~\cite{Malabarba:2020grf} for the details related to the calculation of the integrals in Eq.~(\ref{Int}). Here, we would simply like to state that we are considering an approach in which $\phi(2170)$, $K(1460)$, $f_0(980)$ and $K_1(1270)$ are states generated from hadron interaction dynamics. Thus, a form factor should be associated with each of the three vertices involved in the diagrams in Fig.~\ref{decay}. In this way, when regularizing the $d^3q$ integration in Eq.~(\ref{Int})
\begin{align}
\int d^3q\to (2\pi) \int\limits_0^\infty d|\vec{q}||\vec{q}|^2\int\limits_{-1}^1d\text{cos}\theta\prod\limits_{i=1}^3F_i(|\vec{q}^{\,*}_i|,\Lambda_i),\label{d3q}
\end{align}
where $\theta$ is the angle between the vectors $\vec{q}$ and $\vec{k}$. The index $i=1,\,2,\,3$ in Eq.~(\ref{d3q}), for a given decay process (see Fig.~\ref{decay}), indicates the three vertices involved in the decay mechanism of $\phi(2170)$, $|\vec{q}^{\,*}_i|$ represents the modulus of the center-of-mass modulus momentum related to the vertex $i$ [note that $\vec{q}$  and $\vec{q}^{\,*}$ in Eq.~(\ref{d3q}) are related through a Lorentz boost] and $\Lambda_i$ are as defined in Refs.~\cite{MartinezTorres:2008gy,MartinezTorres:2011gjk,Oller:1997ti,Geng:2006yb} ($\Lambda_{\phi_R\to \phi f_0}\sim 2000$ MeV, $\Lambda_{K_R\to K f_0}\sim 1400$ MeV, $\Lambda_{f_0\to K\bar K}\sim 1000$ MeV, $\Lambda_{K^+_1(1270)\to \phi K^+}\sim 750$ MeV). In Eq.~(\ref{d3q}), $F_i$  is a function representing the form factor considered for the vertex $i$. In case of regularizing the $d^3q$ integral with a sharp cut-off, a Heaviside $\Theta$-function, i.e.,
\begin{align}
F_i=\Theta(|\vec{q}^{\,*}_i|-\Lambda_i),
\end{align}
is used. A monopole form, i.e.,
\begin{align}
F_i=\frac{\bar{\Lambda}^2_i}{\bar{\Lambda}^2_i+|\vec{q}^{\,*}_i|^2},
\end{align}
or an exponential dependence of the type
\begin{align}
F_i=e^{-\frac{|\vec{q}^{\,*}_i|^2}{2\bar{\Lambda}^2_i}},
\end{align}
are also typically used as form factors for the vertices. The value of $\bar{\Lambda}_i$ ($\sim\Lambda_i$) is chosen in such a way that the area under the curve of $F^2_i$ as a function of the modulus of the momentum is same, independently of the form factor used~\cite{Gamermann:2009uq}.

With all these ingredients, we can now determine the decay widths of $\phi(2170)\to K^+(1460) K^-$, $K^+_1(1400) K^-$ and $K^+_1(1270) K^-$. The results obtained are given in Tables~\ref{T1}-\ref{T3}, respectively. As can be seen, consideration of different form factors produces compatible results, thus, the results are basically independent on the regularization procedure.  In case of the decay width of $\phi(2170)\to  K^+(1460) K^-$ (see Table~\ref{T1}) we find a value around $0.8-2.0$ MeV. 

\begin{table}[h!]
\centering
\caption{Partial decay width (in MeV) of $\phi(2170)\to  K^+(1460)K^-$ with different form factors.}\label{T1}
\begin{tabular}{cc}
\hline
Form factor&Decay width\\
\hline\hline
Heaviside-$\Theta$&$1.5\pm0.5$\\
Monopole&$1.3\pm0.4$\\
Exponential&$1.3\pm0.5$\\
\hline
\end{tabular}
\end{table}

In case of the process $\phi(2170)\to K^+_1(1400) K^-$ (see Table~\ref{T2}), the result found for the decay width depends on the model considered to calculate the coupling of $K^+_1(1400)\to \phi K^+$: within model B, which relates  $K_1(1400)$ and $K_1(1270)$ through a mixing angle, the decay width obtained for $\phi(2170)\to K^+_1(1400) K^-$ is around $1.5-3.1$ MeV. 
\begin{table}[h!]
\centering
\caption{Partial decay width (in MeV) of $\phi(2170)\to K^+_1(1400) K^-$ considering the different form factors and the models B and C to describe the properties of $K_1(1400)$.}\label{T2}
\begin{tabular}{ccc}
\hline
Form factor&\multicolumn{2}{c}{Decay width}\\
\hline
&Model B&Model C\\
\hline\hline
Heavise-$\Theta$&$\quad2.6\pm0.5\quad$&$15\pm4$\\
Monopole&$\quad1.9\pm0.4\quad$&$11\pm3$\\
Exponential&$\quad2.1\pm0.4\quad$&$12\pm3$\\
\hline
\end{tabular}
\end{table}
However, if we obtain the $K^+_1(1400)\to \phi K^+$ coupling considering model C, which uses the data from Ref.~\cite{ParticleDataGroup:2020ssz}, the result found for this decay width is $\sim 8-19$ MeV. Such a value represents a sizable contribution of the full width of $\phi(2170)$.  However, it should be mention that the results on the radiative decays in Ref.~\cite{ParticleDataGroup:2020ssz}, and, consequently, the decay width of $\phi(2170)\to  K^+_1(1400) K^-$ found within model C, may need to be taken with caution. This is because the experimental data on the radiative decay of $K^+_1(1270)$ and $K^+_1(1400)$ are obtained, through the Primakoff effect, by assuming them as mixture of states belonging to the axial nonets. Within model A, where $K_1(1270)$ is generated from meson-meson interactions~\cite{Roca:2005nm,Geng:2006yb}, the state $K_1(1400)$ was not found to arise from such dynamics, thus, we cannot calculate the decay width of $\phi(2170)\to K_1(1400)  \bar K$.

In case of the process $\phi(2170)\to K^+_1(1270)K^- $, we find that the decay width (see Table~\ref{T3}) depends on the model used to calculate the coupling of $K^+_1(1270)\to \phi K^+$: within model A, where $K^+_1(1270)$ is considered as a vector-pseudoscalar molecular state with a double pole structure, the decay width obtained is around $1-2$ MeV when taking into account the superposition of the two poles. As can be seen in Table~\ref{T3}, the superposition of the two poles produces non-negligible effects. Note that the mass related to the pole $z_2$ is closer to the value determined from the fit to the experimental data in Ref.~\cite{BESIII:2020vtu}, however, the process $K^+_1(1270) \to \pi K^*(892)$ is considered in Ref.~\cite{BESIII:2020vtu}, where the final state couples rather more strongly to the pole $z_1$. Thus, when comparing our results with the experimental information, it might be more meaningful to consider the decay widths obtained from the superposition of the two poles.
In any case, if the double pole nature of $K_1(1270)$ is confirmed, the results in Ref.~\cite{BESIII:2020vtu} on the related process may require an update. 
\begin{table}[h!]
\caption{Partial decay width (in MeV) of $\phi(2170)\to K^+_1(1270) K^-$ by considering different form factors and the models A, B, C to describe the properties of $K_1(1270)$.\\}\label{T3}
\begin{tabular}{cccccccc}
\hline
Form factor&\multicolumn{7}{c}{Decay width}\\
\hline
&\multicolumn{3}{c}{Model A}&Model B&\multicolumn{3}{c}{Model C}\\
&Poles $z_1$, $z_2$&Pole $z_1$&Pole $z_2$&&Solution $\mathbb{S}_1$& Solution $\mathbb{S}_2$& Solution $\mathbb{S}_3$\\
\hline\hline
Heaviside-$\Theta$&$\quad1.5\pm0.3\quad$&$0.6\pm0.1\quad$&$0.22\pm0.04\quad$&$0.12\pm0.04\quad$&$1.6\pm0.4\quad$&$17\pm3$&$41\pm9$\\
Monopole&$\quad0.8\pm0.2\quad$&$0.3\pm0.1\quad$&$0.12\pm0.02\quad$&$0.07\pm0.02\quad$&$0.9\pm0.2\quad$&$9\pm2$&$23\pm5$\\
Exponential&$\quad1.0\pm0.2\quad$&$0.4\pm0.1\quad$&$0.15\pm0.03\quad$&$0.09\pm0.02\quad$&$1.1\pm0.3\quad$&$11\pm2$&$28\pm6$\\
\hline
\end{tabular}
\end{table}
Next, considering the mixing scheme of model B, the results obtained for the decay width of $\phi(2170)\to K^+_1(1270)K^-$ are similar to those found with model A for the pole $z_2$. Such a result could be in line with the fact that the mass of $K_1(1270)$ in model B is very similar to the mass value associated with the pole $z_2$ in model A. 
In case of the model C, where the experimental data available in Ref.~\cite{ParticleDataGroup:2020ssz} are being used to estimate the couplings of $K^+_1(1270)$ and $K^+_1(1400)$ to the $\phi K^+$ channel, two different scenarios for the decay width of $\phi(2170)\to  K^+_1(1270)K^-$ are found. In one of them (solution $\mathbb{S}_1$), the results are compatible with those obtained in the model A. In the second scenario (solutions $\mathbb{S}_2$ or $\mathbb{S}_3$), a larger decay width for $\phi(2170)\to  K^+_1(1270)K^-$ is obtained, which would constitute a sizable part of the total width of $\phi(2170)$.

After determining the decay widths of $\phi(2170)\to K^-\mathbb{K}^+$, with $\mathbb{K}=K(1460)$, $K_1(1400)$, $K_1(1270)$, we can compare with the experimental results of Ref.~\cite{BESIII:2020vtu}. Note, however, that in the latter reference, the partial decay widths of $\phi(2170)\to K^-\mathbb{K}^+$ were not measured. Instead, the products $\mathcal{B}r\Gamma^{e^+e^-}_R$, with $\Gamma^{e^+e^-}_R$ being the partial decay width of $\phi(2170)\to e^+ e^-$ and $\mathcal{B}r$  the branching fraction for each of the $\phi(2170)\to K^- \mathbb{K}^+$ processes, were extracted. Since the decay width $\Gamma^{e^+e^-}_R$ is not known, we can use the information provided in Ref.~\cite{BESIII:2020vtu} to calculate the ratios
\begin{align}
B_1\equiv\frac{\Gamma_{\phi_R\to K^+(1460)K^-}}{\Gamma_{\phi_R\to K^+_1(1400)K^-}}=\frac{\mathcal{B}r[\phi_R\to K^+(1460)K^-]}{\mathcal{B}r[\phi_R\to K^+_1(1400)K^-]},\label{Br1}
\end{align}
\begin{align}
B_2\equiv\frac{\Gamma_{\phi_R\to K^+(1460) K^-}}{\Gamma_{\phi_R\to K^+_1(1270)K^-}}=\frac{\mathcal{B}r[\phi_R\to K^+(1460) K^-]}{\mathcal{B}r[\phi_R\to K^+_1(1270)K^-]},\label{Br2}
\end{align}
\begin{align}
B_3\equiv\frac{\Gamma_{\phi_R\to K^+_1(1270) K^-}}{\Gamma_{\phi_R\to K^+_1(1400)K^-}}=\frac{\mathcal{B}r[\phi_R\to K^+_1(1270) K^-]}{\mathcal{B}r[\phi_R\to K^+_1(1400)K^-]},\label{Br3}
\end{align}
and compare with our results. Note that although the above ratios do not depend on the coupling $g_{\phi_R\to \phi f_0}$, the particular values found for them are related to the nature, not only of $\phi(2170)$, but also to the one of $K(1460)$, $K^+_1(1270)$ and $K^+_1(1400)$, through the triangular loop mechanisms depicted in Fig.~\ref{decay} and the other vertices involved, which appear as a consequence of considering $\phi(2170)$ as a $\phi f_0(980)$ state. 

In Ref.~\cite{BESIII:2020vtu}, the values (in eV) for the products $\mathcal{B}r\Gamma^{e^+e^-}_R$ are
\begin{align}
\mathcal{B}r[\phi_R\to K^+(1460)K^- ]\Gamma^{e^+e^-}_R&=3.0\pm 3.8,\nonumber\\
\mathcal{B}r[\phi_R\to K^+_1(1400)K^- ]\Gamma^{e^+e^-}_R&=\left\{\begin{array}{c}4.7\pm3.3,~\text{Solution 1}\\98.8\pm7.8,~\text{Solution 2}\end{array}\right.,\nonumber\\
\mathcal{B}r[\phi_R\to  K^+_1(1270)K^-]\Gamma^{e^+e^-}_R&=\left\{\begin{array}{c}7.6\pm3.7,~\text{Solution 1}\\152.6\pm14.2,~\text{Solution 2}\end{array}\right..\label{Brexp}
\end{align}
As we can see in the preceding equation, two possible solutions for $\mathcal{B}r\Gamma^{e^+e^-}_R$ from the fits to the data were obtained in Ref.~\cite{BESIII:2020vtu} in case of the processes $\phi(2170)\to K^+_1(1400)K^- $, $K^+_1(1270)K^- $. Using Eq.~(\ref{Brexp}), we can determine the experimental values for the $B_1$, $B_2$ and $B_3$ ratios of Eq.~(\ref{Br3}), getting
\begin{align}
B^\text{exp}_1&=\left\{\begin{array}{l}0.64\pm0.92,~\text{Solution 1,}\\0.03\pm 0.04,~\text{Solution 2,}\end{array}\right.\nonumber\\
B^\text{exp}_2&=\left\{\begin{array}{l}0.40\pm0.54,~\text{Solution 1,}\\0.02\pm 0.03,~\text{Solution 2,}\end{array}\right.\nonumber\\
B^\text{exp}_3&=\left\{\begin{array}{l}1.62\pm1.38,~\text{Solution 1,}\\1.55\pm 0.19,~\text{Solution 2.}\end{array}\right.\label{Bexp}
\end{align}
Using now the decay widths listed in Tables~\ref{T1}-\ref{T3}, we can determine the ratios in Eqs.~(\ref{Br1}),~(\ref{Br2}),~(\ref{Br3}). The results are given in Tables~\ref{TB1}-\ref{TB3}. Since the decay widths obtained in this work do not depend much on the form factors considered, the values presented for the ratios correspond to the average of the results obtained with different form factors.

\begin{table}[h!]
\centering
\caption{Results for the branching ratio $B_1$. The label ``Experiment'' refers to the values given in Eq.~(\ref{Bexp}).}\label{TB1}
\begin{tabular}{ccc}
\hline
&&$B_1$\\
\hline\hline
\multirow{2}{*}{Our results}&Model B&$0.62\pm0.20$\\
&Model C&$0.11\pm0.04$\\
\hline
\multirow{2}{*}{Experiment}&Solution 1&$0.64\pm0.92$\\
&Solution 2&$0.03\pm0.04$\\
\hline
\end{tabular}
\end{table}
Since the ratio $B_1$ [see Eq.~(\ref{Br1})] involves the decay width of $\phi(2170)\to K^+_1(1400) K^-$, it can be calculated within the models B and C. In the former case, the results obtained are compatible with the experimental value related to solution 1, while in the latter case, the results are closer to the experimental value determined from solution 2. Note that, as a consequence of the uncertainty present in the experimental data, the results obtained within the model C can also be compatible with the value found from solution 1  
\begin{table}[h!]
\centering
\caption{Results for the ratio $B_2$. The label ``Experiment'' refers to the values given in Eq.~(\ref{Bexp}).}\label{TB2}
\begin{tabular}{cccl}
\hline
&&$B_2$\\
\hline\hline
\multirow{7}{*}{Our results}&\multirow{3}{*}{Model A}& $1.3\pm0.4$&(Poles $z_1$, $z_2)$\\
& &$3.6\pm1.2$&(Pole $z_1$)\\
& &$8.8\pm2.8$&(Pole $z_2$)\\
& Model B& $16\pm6$\\
& \multirow{3}{*}{Model C}& $1.2\pm0.4$&(Solution $\mathbb{S}_1$)\\
&  & $0.12\pm0.04$&(Solution $\mathbb{S}_2$)\\
& & $0.05\pm0.02$&(Solution $\mathbb{S}_3$)\\
\hline
\multirow{2}{*}{Experiment}&Solution 1&$0.40\pm0.54$&\\
&Solution 2&$0.02\pm0.03$&\\
\hline
\end{tabular}
\end{table}

The value of $B_2$, as can be seen from Table~\ref{TB2}, depends on the description considered for $K^+_1(1270)$. Within model A [in this case, $K_1(1270)$ has a double pole structure], and considering the interference between the two poles, we obtain a value which is closer to the upper limit for this ratio determined with solution 1 of the BESIII Collaboration. Interestingly, we find that the contribution from the individual poles of $K^+_1(1270)$ produces a larger value for $B_2$, which is not compatible with the experimental value.  Within the model  B, the values determined for $B_2$ are not compatible with those obtained from the experimental data. In case of using model C, solutions $\mathbb{S}_2$ and $\mathbb{S}_3$ produce a value for $B_2$ which is compatible with solution 2 of Ref.~\cite{BESIII:2020vtu} while solution $\mathbb{S}_1$ produces a value compatible with solution 1 of Ref.~\cite{BESIII:2020vtu}. 

\begin{table}[h!]
\centering
\caption{Results for the ratio $B_3$. The label ``Experiment'' refers to the values given in Eq.~(\ref{Bexp}).}\label{TB3}
\begin{tabular}{cccl}
\hline
&&$B_3$\\
\hline\hline
\multirow{4}{*}{Our results}& Model B& $0.04\pm0.01$\\
& \multirow{3}{*}{Model C}& $0.09\pm0.02$&(Solution $\mathbb{S}_1$)\\
&  & $0.96\pm0.16$&(Solution $\mathbb{S}_2$)\\
& & $2.40\pm0.40$&(Solution $\mathbb{S}_3$)\\
\hline
\multirow{2}{*}{Experiment}&Solution 1&$1.62\pm1.38$&\\
&Solution 2&$1.55\pm0.19$&\\
\hline
\end{tabular}\end{table}

In Table~\ref{TB3} we find the results for the ratio $B_3$. Note that this ratio involves the decay width of $\phi(2170)\to K^+_1(1400)K^- $, thus, we can evaluate it within models B and C. Although, because of the similarity between the decay width for $\phi(2170)\to K^+_1(1270) K^-$ within model A (considering the superposition of two poles for $K_1(1270)$) and solution $\mathbb{S}_1$ of model C, it can be inferred that the ratio $B_3$ (under solution $\mathbb{S}_1$ in Table~\ref{TB3}) represent the result for both cases. It can be said, then, that for solution $\mathbb{S}_1$, as well as for model A, the results can be considered to be closer to the lower limit of solution 1 presented in Table~\ref{TB3}. Solutions $\mathbb{S}_2$ and $\mathbb{S}_3$ of model C are compatible with the data. 

In this way, we find that considering $\phi(2170)$ as a $\phi f_0(980)$ molecular states provides a fair description of the ratios $B_1$, $B_2$ and $B_3$ found from the experimental data on $\mathcal{B}r\Gamma^{e^+e^-}_R$ of Ref.~\cite{BESIII:2020vtu}. Further experimental data with higher statistics can be very helpful in drawing more robust conclusions on the properties of $K_1(1270)$ and $K_1(1400)$. The partial decay widths provided in the present work can be useful for future experimental investigations. 

This work is supported by the Funda\c c\~ao de Amparo \`a Pesquisa do Estado de S\~ao Paulo (FAPESP), processos n${}^\circ$ 2019/17149-3, 2019/16924-3 and 2020/00676-8, by the Conselho Nacional de Desenvolvimento Cient\'ifico e Tecnol\'ogico (CNPq), grant  n${}^\circ$ 305526/2019-7 and 303945/2019-2, by the Deutsche Forschungsgemeinschaft (DFG, German Research Foundation), in part through the Collaborative Research Center [The Low-Energy Frontier of the Standard Model, Project No. 204404729-SFB 1044], and in part through the Cluster of Excellence [Precision Physics, Fundamental Interactions, and Structure of Matter] (PRISMA${}^+$ EXC 2118/1) within the German Excellence Strategy (Project ID 39083149) and by the National Natural Science Foundation of China (NSFC), grant n${}^\circ$ 11775099.

\end{document}